\documentclass[osajnl,twocolumn,showpacs,superscriptaddress,11pt]{revtex4-1} 
\usepackage{amsmath,amssymb,graphicx}
\begin{document}

\title{Ultra-low phase noise all-optical microwave generation setup based on commercial devices}

\author{Alexandre Didier}
\author{Jacques Millo}
\author{Serge Grop}
\affiliation{FEMTO-ST Institute, UMR 6174 : CNRS/ENSMM/UFC/UTBM , Time and Frequency Dpt., 26 ch. de l'Epitaphe, 25030 Besan\c{c}on Cedex, France}

\author{Beno\^it Dubois}
\affiliation{FEMTO Engineering, 15 B avenue des Montboucons, 25030 Besan\c{c}on Cedex, France}

\author{Emmanuel Bigler}
\author{Enrico Rubiola}
\author{Cl\'ement Lacro\^ute}\email{Corresponding author: clement.lacroute@femto-st.fr}
\author{Yann Kersal\'e}
\affiliation{FEMTO-ST Institute, UMR 6174 : CNRS/ENSMM/UFC/UTBM , Time and Frequency Dpt., 26 ch. de l'Epitaphe, 25030 Besan\c{c}on Cedex, France}

\begin{abstract}
In this paper, we present a very simple design based on commercial devices for the all-optical generation of ultra-low phase noise microwave signals. A commercial, fibered femtosecond laser is locked to a laser that is stabilized to a commercial ULE Fabry-Perot cavity. The 10 GHz microwave signal extracted from the femtosecond laser output exhibits a single sideband phase noise $\mathcal{L}(f)=-104 \ \mathrm{dBc}/\mathrm{Hz}$ at 1 Hz Fourier frequency, at the level of the best value obtained with such ``microwave photonics'' laboratory experiments \cite{Fortier2011}. Close-to-the-carrier ultra-low phase noise microwave signals will now be available in laboratories outside the frequency metrology field, opening up new possibilities in various domains.
\end{abstract}


\maketitle 

\section{Introduction}
Ultra-low phase noise microwave signals are being used in a growing number of fields. Industrial applications include telecommunication networks, deep-space navigation, high-speed sampling \cite{Valley2007} and radar systems \cite{Scheer1990}. Fundamental physics tests and research experiments also benefit from ultra-stable microwave signals, as in atomic fountain clocks setups \cite{Guena2012}, Lorentz invariance tests \cite{Stanwix2005} or Very Long Baseline Interferometry \cite{Grop2010}.

Such signals are usually generated in three different ways: from a quartz resonator, included in a frequency synthesis \cite{Francois2014, Lautier2014} ; from a sapphire oscillator \cite{Green2006}, often cooled down to cryogenic temperatures \cite{Grop2010a, Hartnett2012} ; or from the optical domain, using an optoelectronic oscillator \cite{Salik2007} or a cavity-stabilized laser and an optical frequency comb \cite{Millo2009b, Fortier2011}. In the latter case, a laser is locked to an ultra-stable Fabry-Perot (FP) cavity, thus providing an ultra-low phase noise optical signal and a  short-term relative frequency stability below $10^{-15}$. This signal is used to phase-lock the repetition rate of an optical frequency comb, which allows for dividing down the signal frequency from the optical to the microwave domain with minimal degradation \cite{Zhang2011}. Progress in the past ten years has allowed to reach extremely low levels of relative frequency stability for FP cavity laser stabilization setups, both by improving the design \cite{Webster2008, Millo2009a, Swallows2012} and materials \cite{Seel1997, Kessler2012} of the cavities. On the other hand, compact and portable FP cavities have been developped for field operation \cite{Leibrandt2011, Webster2011}, and such setups are now commercially available. Fiber-based optical frequency combs have followed the same path and are becoming an essential tool in various experimental physics laboratories.

In this article, we present a setup for all-optical microwave generation based on both a commercial Fabry-Perot cavity and a commercial fibered optical frequency comb. We use this setup to generate an ultra-stable reference signal at 10 GHz, which will later be distributed through the laboratory for future phase-noise characterizations of other oscillators based on Sapphire, Quartz or optical resonators. It will also be complementary to an optical reference signal distributed to French time and frequency laboratories through the REFIMEVE+ network \cite{Refimeve}. In the following sections, we outline the 10 GHz signal generation scheme and analyse the measured signal phase noise and frequency stability.

\section{All-optical microwave generation setup}
\begin{figure}%
\centering
\includegraphics[width=\columnwidth]{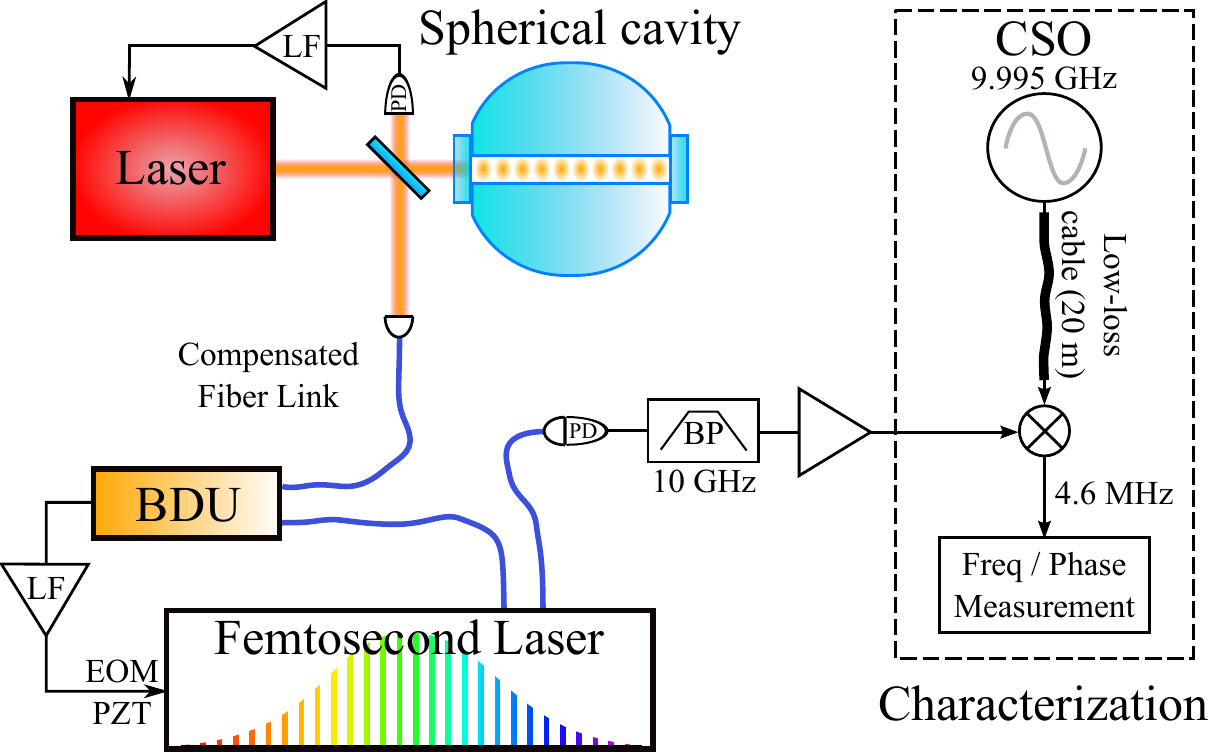}%
\caption{All-optical microwave signal generation and characterization setup. \emph{Generation:} A 1.5 $\mu$m laser is stabilized to a commercial ULE spherical cavity by Pound-Drever-Hall technique. The stabilized output of the laser is used to optically lock a commercial femtosecond laser. The output of the stabilized femtosecond laser is detected by a fibered fast photodiode and filtered and amplified at 10 GHz. \emph{Characterization:} the signal is electronically mixed with the output of a Cryogenic Sapphire Oscillator (CSO) at $9.995 \mathrm{GHz}$. The resulting beatnote at $4.6 \mathrm{MHz}$ is monitored using a frequency counter referenced to a Hydrogen Maser. BDU: beat detection unit; LF: loop filter; PD: photodiode; BP: band pass filter.}%
\label{fig:setup}%
\end{figure}

Our setup for all-optical microwave signal generation is described in Figure \ref{fig:setup}. A commercial continuous-wave (CW) laser at 1542 nm \cite{NKT} is locked to a Fabry-Perot cavity using the Pound-Drever-Hall (PDH) technique. The ultra-stable cavity is a 5 cm long commercial spherical cavity \cite{ATF} based on a design by NIST \cite{Leibrandt2011}. The spherical spacer is held at an optimized angle for minimizing vibration sensitivity \cite{Leibrandt2011}. Fused-Silica mirror substrates are optically contacted to a spherical ULE spacer; ULE rings are placed on the SiO$_2$ substrates to adjust the cavity inversion temperature \cite{Kessler2012a}. The inversion temperature of our cavity was determined to be 10.5$^{\circ}$C, and we measured a finesse of about 400 000 for the fundamental TEM$_{00}$ mode.

We estimate that the thermal noise floor of our cavity will limit the stabilized laser phase noise to $\mathcal{L}(f)=-106 \ \mathrm{dBc/Hz}$ at 1 Hz with a $1/f^3$ slope, corresponding to a relative frequency flicker $\sigma_y\approx 8 \times 10^{-16}$.

The cavity is pumped to ultra-high vacuum using a 2.5 l/s ion pump. The vacuum chamber and the free-space PDH optical setup are placed on a commercial active vibration isolation platform and inside a thermal insulation box, with a total volume of about 0.25 m$^3$. We use homemade electronics for the loop filter, as those are on hand in our laboratory and are usually lower-priced, but similar systems are commercially available. An electro-optical modulator (EOM) modulates the laser phase at $22.5 \ \mathrm{MHz}$ to provide the PDH error signal. The fast corrections are applied to an acousto-optical modulator (AOM) with a bandwidth higher than $100 \ \mathrm{kHz}$, while the slow corrections are applied to the laser's piezoelectric transducer(PZT), with a bandwidth of $50 \ \mathrm{Hz}$. This setup has proven to be very robust, and the laser can stay locked to the FP cavity for weeks without any external intervention.

We optically mix the stabilized laser output with an optical frequency comb produced by a commercial femtosecond laser \cite{Menlo}  using the so-called ``beat detection unit" provided by the manufacturer. This allows for locking the femtosecond laser repetition rate at 250 MHz. A fibered interferometer is readily aligned to detect the beatnote between the optical comb and the reference laser on a high sensitivity photodetector. The output voltage is then processed to generate a lock signal and fed back to an EOM and a PZT placed inside the femtosecond laser cavity to stabilize its repetition rate. The comb carrier envelope offset is stabilized to a radio frequency reference using the so-called $f-2f$ technique \cite{Telle1999}. This is all done using the electronics provided by the manufacturer. Our only addition is a small RF circuit that allows fo the substraction of the CEO to the optical beatnote signal, following Ref. \cite{Zhang2011}. We obtained similar results with and without this substraction scheme.

We detect the 40th harmonic of the repetition rate, near 10 GHz, using a fast photodiode\cite{Discovery}. With an optical power of 3 mW, we obtain about $-30 \ \mathrm{dBm}$ microwave power at 10 GHz. This signal is band-pass filtered at 10 GHz and amplified using two low phase noise microwave amplifiers \cite{Amplis}. The residual phase noise added by such optical division schemes has been measured to be about $-111 \ \mathrm{dBc/Hz}$ at $1 \ \mathrm{Hz}$ with an earlier version of the optical frequency comb \cite{Millo2009}. In principle, this value can even be lowered to $-120 \ \mathrm{dBc/Hz}$ at 1 Hz using additional noise-reduction techniques \cite{Zhang2011}.

The optical fiber link between the ultra-stable laser and the optical frequency comb is actively stabilized using a fiber-noise compensation scheme \cite{Lopez2012}. A fibered AOM is used to correct for optical path fluctuations, with a bandwidth of a few tens of kHz. This can be avoided by placing the FP cavity right next to the femtosecond laser, and using a short optical fiber.

The whole microwave generation setup has stayed locked for days without intervention, even through fairly high temperature fluctuations due to a temporary failure of our air conditionning system. The PDH and the Doppler-cancellation locks have proven to be the most robust, and the femtosecond laser seems to need a quieter acoustic environment. All-in-all, the system is robust enough to continuously provide a 10 GHz ultra-stable reference signal.

\section{Measurements}
The characterization setup of the optically generated microwave signal is illustrated in Figure \ref{fig:setup}. The output of a fast photodiode is filtered at 10 GHz, amplified, and then mixed with a 9.995 GHz signal generated by one of the Cryogenic Sapphire Oscillators (CSO) of the laboratory. The resulting 4.6 MHz beatnote is then sent to a frequency counter referenced to a Hydrogen Maser \cite{Counters}. The CSO has been fully characterized and presents a relative frequency stability below $8 \times 10^{-16}$ for integration times between 1 and 1000 s \cite{Grop2014}. The sapphire whispering gallery mode resonator is held at cryogenic temperature near its inversion point at 6 K, and is integrated in a Pound-Galani oscillator loop. The ultra-stable output is transfered to the ``microwave photonics'' room through a 20 m low-loss coaxial cable without any noise compensation.

\begin{figure}[h!]%
\centering
\includegraphics[width=\columnwidth]{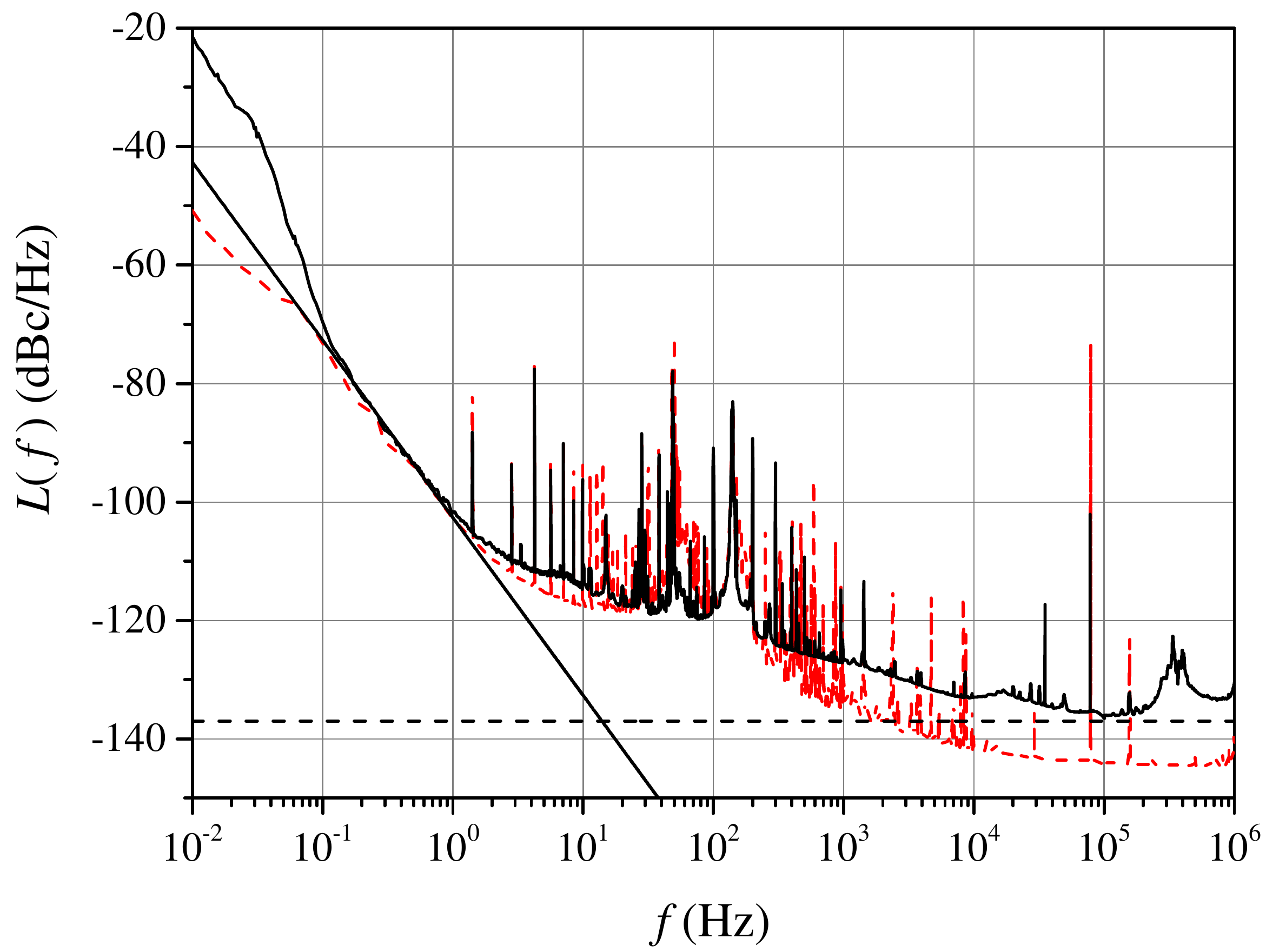}%
\caption{Phase noise of the all-optical microwave signal compared to a CSO signal. \emph{Black curve:} phase noise spectrum of the beatnote between the CSO and the cavity signals. \emph{Black line:} $f^{-3}$ fit of the spectrum between 0.2 and 0.7 Hz. \emph{Red dashed curve:} phase noise spectrum of the beatnote between two identical CSOs. \emph{Dashed line:} photodetection shot noise limit.}%
\label{fig:PSD}%
\end{figure}

We measure a relative phase noise $\mathcal{L}(1 \ \mathrm{Hz})=-102 \ \mathrm{dBc}/\mathrm{Hz}$ for the beatnote, competitive with state-of-the-art optically generated ultrastable microwave signals \cite{Fortier2011, Grop2014}. Figure \ref{fig:PSD} presents the phase noise spectrum. The noise floor is close to the photodetection shot noise limit at $-137 \mathrm{dBc/Hz}$ (dashed line). The spurious peaks between 1 Hz and 100 Hz belong to the CSO phase noise. In particular, resonances at 1.4 Hz and its harmonics are related to the vibrations of the cryo-cooler \cite{Grop2010a}. We plot the phase noise spectrum of the beatnote between two nearly identical CSOs for reference (dashed red line).  It is worth noting that the two measurements do not differ by more than 3 dB between 0.1 Hz and 100 Hz, meaning that the optically generated microwave signal phase noise is very close to the CSO signal phase noise in this frequency range. Moreover, the 20 m coaxial cable might degrade the transfered CSO signal phase noise.

Between 0.2 Hz and 0.7 Hz, the spectrum fits the $f^{-3}$ law (frequency flicker) with a value of $-103 \ \mathrm{dBc/Hz}$ at 1 Hz (black line). This would translate to a relative frequency stability floor of $1.2\times10^{-15}$ for the beatnote. The spectrum shows excess phase noise at frequencies below 0.2 Hz. We believe that this is due to temperature fluctuations in the room, which cause polarization rotations within the PDH optical setup. These rotations induce power fluctuations of the CW laser that couple to the FP cavity resonance frequency.

The phase noise of the CSO we use has been measured to be $\mathcal{L}_{\mathrm{CSO}}(1 \ \mathrm{Hz})=-106 \ \mathrm{dBc/Hz}$. By substracting this value to the beatnote phase noise, we obtain $\mathcal{L}_{\mathrm{opt}}(1 \ \mathrm{Hz})=-104 \ \mathrm{dBc/Hz}$ for the optically generated 10 GHz signal. This is very close to the expected thermal noise floor of the ultra-stable cavity $\mathcal{L}_{\mathrm{cav}}(1 \ \mathrm{Hz})=-106 \ \mathrm{dBc/Hz}$.

\begin{figure}%
\centering
\includegraphics[width=\columnwidth]{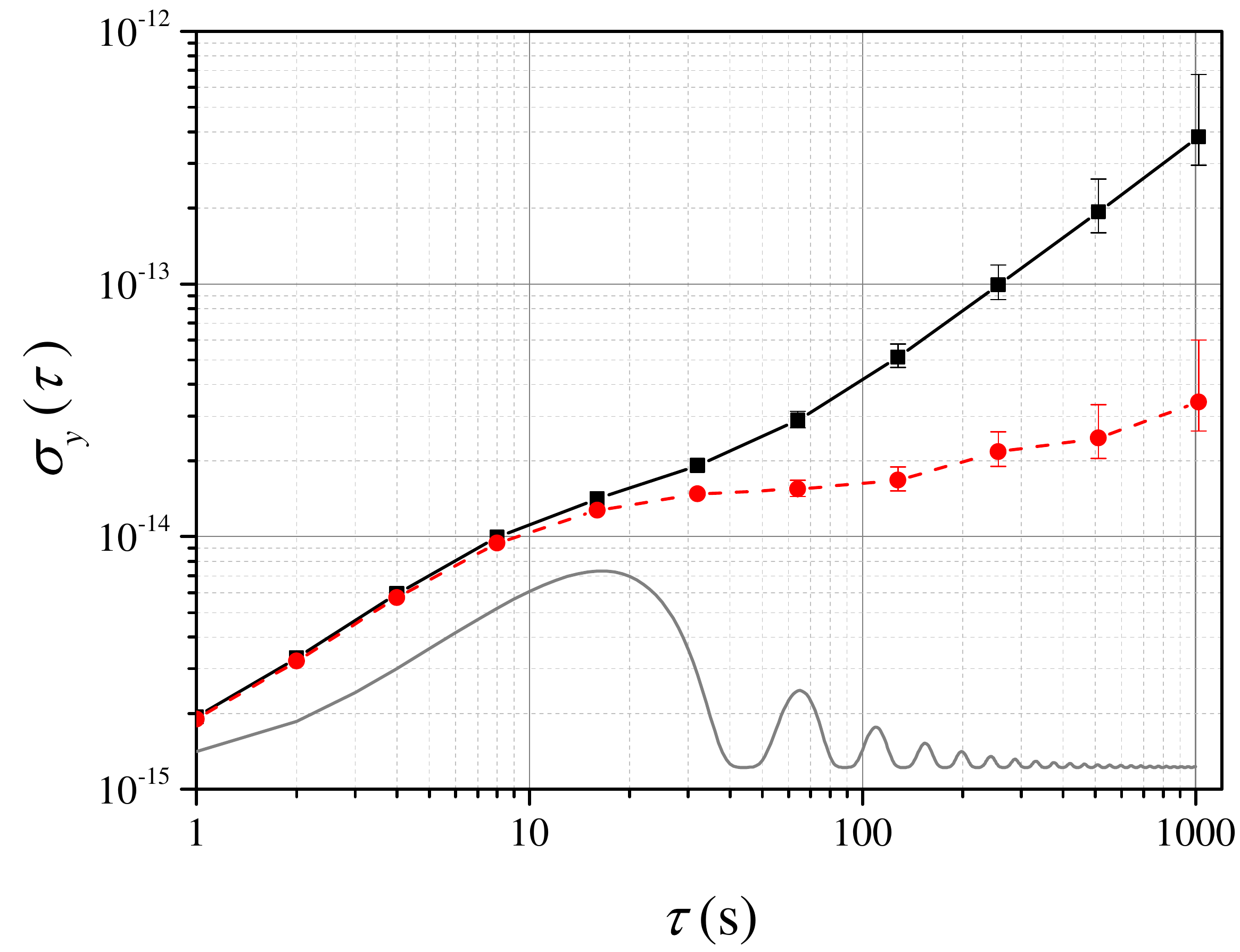}%
\caption{Allan deviation of the all-optical microwave signal compared to a CSO signal. \emph{Red dashed curve:} relative frequency stability with linear drift removed. \emph{Light gray curve:} estimated contribution of the $26 \ \mathrm{mHz}$ modulation to the Allan deviation.}%
\label{fig:stab}%
\end{figure}

Figure \ref{fig:stab} presents the  relative frequency stability of the optically generated microwave signal versus the CSO. We obtain $\sigma_y(1\mathrm{s})=1.9\times10^{-15}$ for the beatnote, higher than the flicker frequency floor ($1.2 \times 10^{-15}$). This is mostly due to excess frequency noise at low frequencies. In particular, a parasitic modulation of the beatnote at $26 \ \mathrm{mHz}$ (most likely due to room-temperature fluctuations) degrades the signal relative frequency stability between 1 and 20 seconds. We have numerically extracted the relative frequency power spectral density $S_y$ at this frequency from the drift-removed temporal dataset. We plot the relative frequency stability obtained with such a purely sinusoidal modulation added to the Flicker floor at $1.2\times 10^{-15}$  for reference (gray line - see \cite{Kersale2006} for details). The initial slope and frequency stability fairly agrees with our measurement.

The linear drift of the frequency leads to a $3.8 \times 10^{-16} \ \tau$ stability for integration time longer than 200 s. Potential improvements of the short-term relative frequency stability include the better rejection of the room-temperature fluctuations, as well as a refined measurement of the inversion temperature of the cavity.

\section{Conclusion}
In summary, we have presented the first all-optical setup for microwave signal generation based on commercially available instruments. This setup shows a phase noise spectrum competitive with the best reported values both for all-optical setups \cite{Fortier2011} and cryogenic sapphire oscillators \cite{Grop2010a}.

To this day, such ``microwave photonics'' setups are still found mostly in metrology institutes, as they used to require the design of an ultra-stable FP cavity and/or optical frequency comb. The setup that we present in this article should allow the spreading of optical microwave generation outside of frequency metrology labs, thanks to the availability of the key-devices and the overall simplicity of the setup. This will pave the way to tantalizing new developments in fields such as high-resolution spectroscopy, atomic physics and very-long baseline interferometry.

\section{Acknowledgements}
The authors would like to thank Rodolphe Boudot and Vincent Giordano for their careful reading of the manuscript as well as Christophe Fluhr for fruitful discussions about the CSO and Maser performances.

This work is funded by the Fond Europ\'een de d\'eveloppement R\'egional (FEDER).
This work is also funded by the ANR Programme d'Investissement d'Avenir (PIA) under the Oscillator IMP project and First-TF network, and by grants from the R\'egion Franche Comt\'e intended to support the PIA.

\bibliographystyle{osajnl}

\begin{thebibliography}{99}
\bibitem{Counters}
We have used a KnK counter to obtain the stability curve presented in this
  manuscript, and a Symmetricom 5125A to acquire the phase noise traces. Both
  counters give the same result for stability measurements.
\end{thebibliography}

\end{document}